\documentclass[aps,preprint,showpacs]{revtex4}
\begin{document}        %
\draft
\title{Comments on "Slip coefficient in nanoscale pore flow"}
\author{Zotin K.-H.  Chu} 
\affiliation{3/F, 24, 260th. Lane, First Section, Muzha Road,
Wen-Shan District, Taipei, Taiwan 116,  China}
\begin{abstract}
We make some remarks on  Sokhan and Quirke's [{\it Phys. Rev. E}
78, 015301(R) (2008)] paper. Sokhan and Quirke mentioned that,
considering their main result, {the slip coefficient is
independent of the external force (flux)} which is not consistent
with previous measurements and approaches. We also discuss the
sudden changes of the slip coefficient for larger Knudsen numbers
or smaller nanopores.
%
\end{abstract}
\maketitle
\bibliographystyle{plain}
Sokhan and Quirke just presented an expression for the slip
coefficient ($l_s$) applicable to flows in nanoscale pores, which,
as they claimed [1], has been verified by nonequilibrium
molecular-dynamics simulation. Their results showed that the slip
coefficient depends strongly on the pore width for small pores
tending to a constant value for pores of width ($H$) $>$ 20
molecular diameters ($\sigma$) for their systems, in contrast to
the linear scaling predicted by Maxwell's theory of slip. Note
that ${\bf u}_s= l_s \nabla {\bf u}$ in [1] is of doubt as the
left-hand-side (the slip velocity) is a vector while the
right-hand-side is generally a tensor ($\nabla {\bf u}$ : the
shear rate). To the best knowledge of the present author, the slip
velocity is normally defined in one-dimensional sense
[2-3].\newline
The present author, however, based on his previous experiences [2]
(the gravity-driven case in [1] is similar to that of electric
field-driven case in [2] considering the external force), would
like to make some remarks about their presentation [1]. To begin
with, we like to recall the recent critical review : '{\it The
continuum approximation seems to break down below $10$ nm in case
of water in nanotubes. Smooth liquid-gas interface (meniscus)
disappears in tubes with the diameter less than $8$-$10$ nm and an
anomalous behavior of water is observed in $1$-$7$ nm carbon
nanotubes}' [4]. The main results for a {\it nanoscale} pore flow
claimed in [1] (cf. Eq. (8) therein) :
\begin{equation}
 l_s =\frac{\tau \eta}{\rho h}-\frac{H}{6},
\end{equation}
where $\tau$ is the relaxation time, $\eta$ : the interfacial
shear viscosity, $\rho$ : mass density of the fluid and $h=H/2$,
by Sokhan and Quirke are still mainly based on the continuum
approximation : Equations (2),(3),(6) in [1] which contradict to
the summarized results just mentioned above [4] and other
researchers' claim [5]. There is no wonder that the authors of [1]
made remarks : {\it Deviation of the order of 10\% from the bulk
values for the $H=5 \sigma$ (cf. the detailed explanation of the
notations in [1]) pore is due to the overlap of adsorbed layers at
two surfaces and to inaccuracy in determination of the reference
bulk density. At lower densities, when Kn $\ge 1$, the viscosity
in the pore deviates markedly from bulk values}. To be concise,
the continuum approximation breaks down for Kn $\ge 1$ [2,6]. On
the other hand, the present author argues that Sokhan and Quirke
should illustrate the realistic range of $H$ and $\sigma$ in terms
of dimensional units for their calculations in [1]. What are the
exactly valid minimum dimensions for $H$ and $\sigma$ considering
their numerical approaches [1]? Note that Mattia and Gogotsi
raised the question : Whether the nonslip boundary condition is
applicable at very small scales remains open [4].
\newline Secondly, it was mentioned in [1] (cf. statements near Eq. (8) in
[1]), saying that the main result of [1],
 that the slip coefficient
(or slip length) is independent of the external force (flux), but
depends nonlinearly on the pore width, both directly and
indirectly through the relaxation time (cf. Eq. (1) above). This
result is quite different from those presented by Majumder {\it et
al.} [7] (cf. Table 1 in [7] especially for those different
external-forcing measurements or calculations for the slip length
about water). At least, considering Majumder {\it et al.}'s data
[7], for the same fluid (water) : increasing the external forcing
will increase the slip coefficient (or slip length). Meanwhile
this result is also different from those presented in [2] or [8]
where the external forcing can adjust the (averaged) velocity
profiles (not to mention the slip velocity near the walls or
interfaces). To be specific, the present author likes to argue
that there is a possibility that {\it grazing} collisions between
particles (atoms and molecules) and surfaces (or boundaries) [9]
occurred for those particles approaching to as well as reflecting
from or along the nanopore walls. The latter could be neglected by
Sokhan and Quirke in [1]. In essence, the grazing collisions
between particles and boundaries are mainly tuned by the external
forcing [9].
\newline
Thirdly,  Sokhan and Quirke claimed [1] for the systems with the
same Knudsen number (Kn=$\lambda$ (mean free path)/$H$ (pore
width)), the slip coefficient increases with the pore width (cf.
Fig. 2 in [1]). This might be trivial in microdomains (but not in
nanodomains) from previous kinetic theory as the Knudsen number
being fixed (once $H$ increases then $\lambda$ should also
increase so that Kn is kept to be the same; considering Sokhan and
Quirke's derivation :
\begin{equation}
 l_s =\lambda \frac{2}{\alpha}-\frac{H}{6},
\end{equation}
compared to the Maxwell's relation :
\begin{equation}
 l_s =\lambda(\frac{2}{\alpha}-1)
\end{equation}
cf. Eqs. (9) and (1) in [1]; here $\alpha$ defines a fraction of
specularly reflected molecules or $\alpha$ defines the fraction of
the flux of tangential momentum transmitted in collisions and is
called the 'accommodation coefficient' (TMAC) [1]) then in
transitional flow regime (Kn $\approx$ O(1)) due to the increasing
of the mean free path, the slip velocity (proportional to the slip
coefficient or slip length) will also increase (cf. Fig. 3 in
[2]). On the other hand, to check the trend illustrated in Fig. 3
of [1], we have, using Eq. (9) in [1] and $\rho=0.125 m/\sigma^3$
(from $n^*=0.125=\rho\sigma^3/m$, please see the first line  of
the right column  at page 4 of [1]),
\begin{equation}
 l_s =\frac{32 \eta \sigma^3}{m \,\bar{c}\,
 \alpha}-\frac{H}{6}\equiv l_s|_{SQ}
\end{equation}
which is related to the main result derived by Sokhan and Quirke
in [1] and
\begin{equation}
 l_s =\frac{32 \eta \sigma^3}{m \,\bar{c}\, \alpha}-\frac{16 \eta \sigma^3}{m
 \bar{c}} \equiv l_s|_{Max}
\end{equation}
considering the Maxwell's expression (i.e., Eq. (1)) in [1], where
$\bar{c}$ is the mean speed of the molecules. The difference of
above two expressions then give us the understanding of those
shown in Fig. 3 of [1]:
\begin{equation}
l_s|_{Max}- l_s|_{SQ}=\frac{16 \eta \sigma^3}{m
 \bar{c}}-\frac{H}{6} \equiv\lambda-\frac{H}{6}.
\end{equation}
Unfortunately, even we normalized this difference with respect to
$\sigma$ (the same as that of Fig. 3 of [1]), the obtained
difference is still dependent upon $\eta$, $m$, and $\bar{c}$ (or
$\lambda$). However, we cannot get the values of $\eta$, $m$, and
$\bar{c}$ (or $\lambda$) from [1] to evaluate the further detailed
difference. Can Sokhan and Quirke clarify this issue? What is the
Kn for calculations illustrated in Fig. 3 of [1]? Meanwhile for $H
\sim \sigma$, what is the realistic {\it mean free path} for
particles or molecules in confined nanodomains? This is crucial to
our judgement whether Fig. 3 in [1] is universal?
\newline
Finally, the present author, based on the sudden decrease of the
slip coefficient of case $H=5 \sigma$ (w.r.t. $H=10 \sigma$) for
the larger Kn (say, Kn $>3.0$) in Fig. 2 of [1], also wonders that
once $H < 5 \sigma$, the curves of $l_s/H$ vs. Kn will become
nonmonotonic as the decreasing of $l_s/H$ for smaller $H$ (less
than $5 \sigma$) will continue? We argue that the possible cause
is due to the  estimation from the relaxation times and bulk
viscosities at densities equal to that in the middle of the pore
considering Sokhan and Quirke's calculations [1]. It seems to us
the latter treatment should be refined in much more narrow
nanopores considering the other realistic physical influences due
to confinement (e.g., is the mean free path definition still be
valid in confined nanodomains?).

\end{document}